\magnification=1200
\hsize=15truecm
\vsize=23truecm
\baselineskip 18truept
\voffset=-0.5 truecm
\parindent=0.5cm
\overfullrule=0pt
\def\R{{\rm R}}
\def\F{{\rm F}}
\font\titolo=cmbx10 scaled\magstep2

\catcode`@=11
%
%
%
\def\lsim{\mathchoice
  {\mathrel{\lower.8ex\hbox{$\displaystyle\buildrel<\over\sim$}}}
  {\mathrel{\lower.8ex\hbox{$\textstyle\buildrel<\over\sim$}}}
  {\mathrel{\lower.8ex\hbox{$\scriptstyle\buildrel<\over\sim$}}}
  {\mathrel{\lower.8ex\hbox{$\scriptscriptstyle\buildrel<\over\sim$}}} }
\def\gsim{\mathchoice
  {\mathrel{\lower.8ex\hbox{$\displaystyle\buildrel>\over\sim$}}}
  {\mathrel{\lower.8ex\hbox{$\textstyle\buildrel>\over\sim$}}}
  {\mathrel{\lower.8ex\hbox{$\scriptstyle\buildrel>\over\sim$}}}
  {\mathrel{\lower.8ex\hbox{$\scriptscriptstyle\buildrel>\over\sim$}}} }
\def\croce{\displaystyle / \kern-0.2truecm\hbox{$\backslash$}}
\def\lqua{\lower4pt\hbox{\kern5pt\hbox{$\sim$}}\raise1pt
\hbox{\kern-8pt\hbox{$<$}}~}
\def\gqua{\lower4pt\hbox{\kern5pt\hbox{$\sim$}}\raise1pt
\hbox{\kern-8pt\hbox{$>$}}~}
\def\mma{\lower1pt\hbox{\kern5pt\hbox{$\scriptstyle <$}}\raise2pt
\hbox{\kern-7pt\hbox{$\scriptstyle >$}}~}
\def\mmb{\lower1pt\hbox{\kern5pt\hbox{$\scriptstyle >$}}\raise2pt
\hbox{\kern-7pt\hbox{$\scriptstyle <$}}~}
\def\mmc{\lower4pt\hbox{\kern5pt\hbox{$<$}}\raise1pt
\hbox{\kern-8pt\hbox{$>$}}~}
\def\mmd{\lower4pt\hbox{\kern5pt\hbox{$>$}}\raise1pt
\hbox{\kern-8pt\hbox{$<$}}~}
\def\lsu{\raise4pt\hbox{\kern5pt\hbox{$\sim$}}\lower1pt
\hbox{\kern-8pt\hbox{$<$}}~}
\def\gsu{\raise4pt\hbox{\kern5pt\hbox{$\sim$}}\lower1pt
\hbox{\kern-8pt\hbox{$>$}}~}
\def\croce{\displaystyle / \kern-0.2truecm\hbox{$\backslash$}}
\def\ali{\hbox{A \kern-.9em\raise1.7ex\hbox{$\scriptstyle \circ$}}}
\def\2frecce{\hbox{\lower 0.3ex\hbox{$\leftarrow$} 
\hbox{\kern-1.3em\raise 0.3ex\hbox{$\rightarrow$}}}}
%
%
%
%
\def\quad@rato#1#2{{\vcenter{\vbox{
        \hrule height#2pt
        \hbox{\vrule width#2pt height#1pt \kern#1pt \vrule width#2pt}
        \hrule height#2pt} }}}
\def\quadratello{\mathchoice
\quad@rato5{.5}\quad@rato5{.5}\quad@rato{3.5}{.35}\quad@rato{2.5}{.25} }
%
%
\font\s@=cmss10\font\s@b=cmbx8
\def\reali{{\hbox{\s@ l\kern-.5mm R}}}
\def\m{{\hbox{\s@ l\kern-.5mm M}}}
\def\k{{\hbox{\s@ l\kern-.5mm K}}}
\def\naturali{{\hbox{\s@ l\kern-.5mm N}}}
\def\interi{{\mathchoice
 {\hbox{\s@ Z\kern-1.5mm Z}}
 {\hbox{\s@ Z\kern-1.5mm Z}}
 {\hbox{{\s@b Z\kern-1.2mm Z}}}
 {\hbox{{\s@b Z\kern-1.2mm Z}}}  }}
\def\complessi{{\hbox{\s@ C\kern-1.7mm\raise.4mm\hbox{\s@b l}\kern.8mm}}}
\def\toro{{\hbox{\s@ T\kern-1.9mm T}}}
\def\unity{{\hbox{\s@ 1\kern-.8mm l}}}
%
%
\font\bold@mit=cmmi10
\def\setbmit{\textfont1=\bold@mit}
\def\bmit#1{\hbox{\textfont1=\bold@mit$#1$}}
%
\catcode`@=12

\null
\vskip -1truecm
\rightline{DFPD/96/TH/07}
\rightline{February 1996}
\vskip 0.5truecm

\centerline
{\titolo Worldvolume and target space anomalies in the}
\centerline
{\titolo D=10 super--fivebrane sigma--model${}^*$}

\vskip 1.5truecm
\centerline
{\bf K. Lechner and M. Tonin}

\vskip 1truecm

\centerline
{\it Dipartimento di Fisica, Universit\`a  di Padova}

\vskip 0.3truecm

\centerline
{\it and}

\vskip 0.3truecm

\centerline
{\it Istituto Nazionale di Fisica Nucleare, Sezione di Padova}
\centerline
{\it Italy}

\vskip 1.5truecm

\centerline
{\bf Abstract} 

\vskip 0.3truecm

The fields of the conjectured ``heterotic" super--fivebrane sigma--model in
ten dimensions are made out of a well known gravitational sector, the $X$
and the $\vartheta$, and of a still unknown heterotic sector which should
be coupled to the Yang--Mills fields. We compute the one--loop $d=6$
worldvolume and $D=10$ target space Lorentz--anomalies which arise from the
gravitational sector of the heterotic super--fivebrane sigma--model, using
a method which we developed previously for the Green--Schwarz heterotic
superstring. These anomalies turn out to carry an overall coefficient which
is $1/2$ of that required by the string/fivebrane duality conjecture. As a
consequence the worldvolume anomaly vanishes if the heterotic fields
consist of 16 (rather than 32) complex Weyl fermions on the worldvolume. 
This implies that the string/fivebrane duality conjecture
can not be based on a ``heterotic"
super--fivebrane sigma--model with only fermions in the heterotic sector.
Possible implications of this result are discussed. 

\vskip 0.5truecm

\noindent
$^*$ Supported in part by M.P.I. This work is carried out in the framework of
the European Community Programme Gauge Theories, Applied 
Supersymmetry and Quantum Gravity" with a financial contribution under 
SC1--CT--92--D789.

\vfill\eject

\noindent
{\bf 1. Introduction}

\vskip 0.5truecm

It is known from longtime that $N=1$ Supergravity in ten dimensions 
exists in two variants: the $B_2$--version $^{[1]}$ which involves a 
two--form $B_{mn}$, naturally coupled to superstrings, and the 
$B_6$--version $^{[2]}$ which involves a six--form $B_{m_1...m_6}$, naturally 
coupled to super--fivebranes. The two versions are dual to each other in 
the sense that the field strength $H_3$ of $B_2$ is the dual of the field 
strength $H_7$ of $B_6$. When these Supergravities are coupled to a
Super--Yang--Mills theory
the Green--Schwarz anomaly cancellation mechanism $^{[3]}$ 
works in both cases $^{[3,4]}$ provided the gauge group is  
$E_8 \otimes E_8$ or $SO(32)$ and the field strengths $H_3$ 
or $H_7$ are modified by 
suitable Chern--Simons terms.
More recently it has been discovered that the $B_2$ Supergravity admits non
singular solitonic fivebrane solutions $^{[5],[6]}$ and, viceversa, the $B_6$ 
Supergravity admits non singular heterotic string solutions $^{[7]}$.

These results led to the conjecture that the heterotic string and the 
``heterotic" fivebrane in ten dimensions are dual to each other, meaning 
that the strong coupling regime of the string is described by the weak 
coupling regime of the fivebrane and viceversa $^{[5-14]}$. This 
implies that quantum loop effects of the string should 
correspond to sigma--model loop effects (tree level) of the fivebrane.
The strong/weak coupling duality between heterotic strings and fivebranes
in $D=10$ is supported also by the fact that their 
low energy bosonic effective actions are related 
by a rescaling of the metric which gives rise to an inversion of the
quantum loop expansion parameter.

On the other hand this duality conjecture presents also some problematic 
aspects, see e.g. refs. [15,16],
and it is, moreover, difficult to test since a consistent formulation of 
the ``heterotic" super fivebrane is still lacking: the classical 
$\kappa$--invariant action for the gravitational sector of the 
$D=10$ super--fivebrane is well known$^{[17]}$, 
but a $\kappa$--invariant action for the heterotic sector, 
which couples to the $D=10$ target Super--Yang--Mills fields, 
is not yet known. Indeed, it appears a difficult if not impossible
task to find it since every simply minded action would destroy 
$\kappa$--invariance.

In spite of these difficulties,
it is, however, clear that fivebranes are fated to play a role in 
string duality and that it is worthwhile to analyse
``heterotic" super--fivebrane models with the ultimate goal
of finding a consistent formulation for them. An important step in this
direction, and this is the purpose of the present paper,
would be constituted by the knowledge of their
worldvolume and target space anomalies: the formers are 
analogous to the worldsheet conformal anomalies (central charge)
in the string and have to cancel in a consistent model; actually, the
requirement of their cancellation constrains the field
content of the heterotic sector for the string and for the fivebrane.
The latters are ``genuine sigma--model" effects and determine heavily
the structure of the $N=1,D=10$ supergravity theory, in which the
sigma--model is embedded, in that they 
should cancel via the (dual of) the Green--Schwarz mechanism.

A first attempt in this direction has been performed in [13]. Here,
with a somehow conjectural calculation which has, however, been 
questioned in [18], the target space 
anomaly polynomial for the $D=10$, ``heterotic" fivebrane with 
gauge group $SO(32)$ has been evaluated and found to be in agreement with the 
string/fivebrane duality conjecture if one assumes that the heterotic
sector is made out of 32 complex $d=6$ Weyl fermions.

Recently we performed  a systematic analysis $^{[19,20]}$ of the one--loop 
anomalies of the Green--Schwarz heterotic string sigma--model. In 
ref. [20] we determined, in particular,
the worldsheet ``genuine string"  
anomaly (i.e. the one which survives in the flat limit) in a covariant 
background gauge, generalizing a method 
first proposed by Wiegmann $^{[21]}$, and we have shown that it vanishes in 
ten dimensions.

In this paper we apply the methods used in $^{[19,20]}$ to compute the 
one--loop target space and worldvolume Lorentz anomalies stemming from the
gravitational sector of the $D=10$, 
super--fivebrane sigma--model 
as described by the $\kappa$--invariant action of 
ref. [17]. We perform the calculation in the framework of the background 
field method combined with a normal coordinate expansion $^{[22]}$. This 
allows us to keep the invariance under target space $SO(1,9)$ 
transformations manifest at the classical level and to impose a covariant 
background gauge to fix $\kappa$--symmetry.

We compute first the target space Lorentz anomaly following the approach
of ref. [19]. This anomaly receives contributions only from the 
functional integration
over the quantum counterparts, $y^\alpha$, of the chiral fermionic fivebrane 
$\vartheta$--fields. 
Once this anomaly is known, we can perform a finite target space local $SO(1,9)$
rotation, as in [20], to transform the kinetic term for the
$y'$s, which depends also on the target fields,
to a canonical kinetic term supplemented by a Wess--Zumino action. At this 
point it becomes rather straightforward to compute the total 
worldvolume Lorentz anomaly arising from the gravitational 
sector of the fivebrane.

Since the field content and the corresponding action of the heterotic fivebrane
sector is not yet known,
the target space (gauge) and worldvolume anomalies coming from this sector
can not be computed from first principles.
However, by assuming that this sector is made out of  $N_\psi$ 
complex $d=6$ Weyl fermions 
coupled to the target space gauge fields and the 
worldvolume metric (as assumed also in the canonical string/fivebrane
duality conjecture$^{[13]}$) these anomalies  can 
be computed via the index theorem even in the absence of an explicit  form
of the action. Therefore, under this assumption, we get an explicit
expression of the total worldvolume and target space
anomaly polynomial for the would--be heterotic 
fivebrane, see formula (34).

It happens that the worldvolume anomaly cancels if $N_\psi=16$ meaning that
the heterotic fermions have to be 16 rather then the 32 which are expected 
in the canonical duality conjecture. On the 
other hand, the remaining purely 
target space anomaly polynomial, which should be cancelled via
the dual Green--Schwarz mechanism, does not match the polynomial predicted
by duality in that the relative coefficients of the terms 
$tr\R^4$ and $tr\F^4$ differ by a factor of 1/2. This is
due to the fact that the target space Lorentz anomaly
arising from the gravitational fivebrane sector, which 
is triggered essentially by the term $tr\R^4$,  turns out to be 1/2 of that
usually expected. The possible meaning of 
these results will be discussed at the end of the paper.

\vskip 0.5truecm

\noindent
{\bf 2. The Super--fivebrane $\sigma$--model}

\vskip 0.5truecm

The action for the gravitational sector
of the super--fivebrane sigma--model embedded in an $N=1$, $D=10$
target space supergravity background$^{[17]}$ is given by 
$$
S_6=-{1\over(2\pi)^3\beta^\prime} \int d^6\sigma
\left( {1\over 2} \sqrt{g} g^{ij} v_i^a v_{ja} - {1\over 6!} 
\ \varepsilon^{j_1\cdots j_6} V_{j_1}^{A_1}\cdots V_{j_6}^{A_6} 
B_{A_6 \cdots A_1} -
2 \sqrt{g}\right).\eqno(1)
$$
The fivebrane fields are the supercoordinates 
$Z^M=(X^m(\sigma),\vartheta^\mu(\sigma))$, $(m=0,1,...,9$ and
$\mu=1,...,16)$ and the worldvolume metric $g^{ij}(\sigma)$ where
$i,j=0,1,...,5$ are curved worldvolume indices. In what follows it 
will be useful 
to write the metric in terms of sechsbeins, 
$g^{ij} = e^i_{\widehat a}\ 
e^j_{\widehat b} \ \eta^{\widehat a \ \widehat b}$, where 
$\widehat a, \widehat b=0,1,...,5$ are
flat $SO(1,5)$ indices.
We set $V_i{}^A(Z)=\partial_i Z^M E_M{}^A(Z)$ and $v_i{}^A=
e^{-1/3\varphi} V_i{}^A$ where $E_M{}^A$ is the target space 
superzehnbein, $\varphi$ is the dilaton superfield and the flat $SO(1,9)$ index
$A$ stands for ten bosonic and sixteen fermionic entries,
$A=(a,\alpha)$ $(a=0,1,...,9$ and $\alpha=1,...,16)$. The
$D=10$ six--superform $B_6 = {1\over 6!} E^{A_1}\cdots E^{A_6} 
B_{A_6\cdots A_1}$,
where $E^A=dZ^M E_M{}^A$, appears in (1) through its pullback on the 
$d=6$ worldvolume of the fivebrane. 

The symmetries of the action are given by 
the $d=6$ diffeomorphisms with parameter $c^j$,  
$\kappa$--transformations with parameter
$\kappa^\alpha$ (a $D=10$ spinor and $d=6$ scalar) and, if the metric is 
replaced by the sechsbeins, local worldvolume $SO(1,5)$ Lorentz 
transformations with
parameter $\ell_{\widehat a \ \widehat b}$:
$$
\eqalign
{
\delta Z^M & = \Delta^\alpha E_\alpha{}^M + c^j \partial_j Z^M\cr
\delta g^{ij} & = 2X^{ij} - {1\over 2} \ g^{ij} X^{hk} g_{hk} + 
c^k \partial_k g^{ij} - 2\partial_k c^{(i}g^{j)k}\cr
\delta e^i_{\widehat a} &= \ell_{\widehat a}{}^{\widehat b} e^i_{\widehat b} + c^j 
\partial_j e_{\widehat a}^i-\partial_j c^i e_{\widehat a}^j+
(\kappa-{\rm transformations}).\cr
}
\eqno(2)
$$

\smallskip

\noindent
We have set:
$$
\Delta^\alpha = (\unity + \Gamma)^\alpha{}_\beta \kappa^\beta\eqno(3)
$$
$$
\Gamma^\alpha{}_\beta = {1\over 6!\sqrt{g}}\ \varepsilon^{j_1\cdots j_6} 
v_{j_1}{}^{a_1}\cdots v_{j_6}{}^{a_6}
(\Gamma_{a_1...a_6})^\alpha{}_\beta\eqno(4)
$$
\noindent
and $X^{ij}$ is given in the appendix.

Actually, $\kappa$--invariance of the action is achieved once the target 
space superforms satisfy suitable constraints. Introducing the $SO(1,9)$ 
superconnection $\Omega_A{}^B=E^C\Omega_{CA}{}^B, \
\Omega_\alpha{}^\beta =
{1\over 4} (\Gamma_{ab})_\alpha{}^\beta\Omega^{ab},\ 
\Omega_a{}^\alpha=\Omega_\alpha
{}^a=0$, $\Omega^{ab}=-\Omega^{ba}$, 
the supertorsion $T^A=dE^A+E^B \Omega_B{}^A ={1\over 2} E^B E^C 
T_{CB}{}^A$ and the $B_6$--supercurvature $H_7=dB_6$, following the
conventions of ref. [27], these constraints read
$$
\eqalign
{
T_{\alpha\beta}{}^a & = 2 (\Gamma^a)_{\alpha\beta}
\cr
T_{\alpha a}{}^b &=0 
\cr
H_{\alpha\beta a_1...a_5}  &= - 2e^{-2\varphi}
(\Gamma_{a_1...a_5})_{\alpha\beta}
\cr
H_{\alpha a_1...a_6}& =-2e^{-2\varphi} (\Gamma_{a_1...a_6})_\alpha{}^\beta
D_\beta\varphi,
\cr
}
\eqno(5)
$$
and the components of $H_7$ with  more than two spinorial indices are zero.

The action (1) is also invariant under  $SO(1,9)$ ``external" local Lorentz
transformations, with parameter $L_A{}^B, L_\alpha{}^\beta = {1\over 4}
(\Gamma_{ab})_\alpha{}^\beta L^{ab},\
L_a{}^\alpha=L_\alpha{}^a=0
\  L_{ab}=-L_{ba}$, under which we have
$$\eqalign{
\delta_L \Omega_A{}^B & = dL_A{}^B + L_A{}^C \Omega_C{}^B -
\Omega_A{}^C L_C{}^B\cr
\delta_L E^A & = - E^B L_B{}^A.\cr}\eqno(6)
$$
To determine the worldvolume anomalies we choose to keep the effective 
action invariant under $d=6$ diffeomorphisms, at the expense of  local 
$SO(1,5)$ Lorentz anomalies, and we trigger the ``genuine sigma--model" 
$\kappa$--anomalies (the ones which go to zero when the target fields are 
switched off) through the $SO(1,9)$ local Lorentz--anomalies; the latters 
are tied to the formers by a coupled cohomology problem (see the 
introduction in ref. [19]).

We apply the background field method supplemented by a normal coordinate
expansion to
keep the classical action manifestly $SO(1,9)$ invariant. So we write
$Z=Z_0+\Pi(Z_0,y)$, treat the $Z_0$ as classical fields and perform the 
functional integration over the (flat) quantum variables $y^A=(y^a, 
y^\alpha)$. The worldvolume sechsbein $e^i_{\widehat a}$ is treated as purely 
classical and constrained to satisfy the classical equations of motion 
$$
v_i{}^a v_{ja} = g_{ij}.\eqno(7)
$$
Here, and in what follows, the $v_i^a$ and all target space fields 
are evaluated at $Z_0$.

Since (non trivial) local Lorentz anomalies can arise only from the 
integration over the fermionic $y^\alpha$ we concentrate now on these 
variables only. First of all, the constraint (7) allows to perform an 
$SO(1,9)$ covariant $\kappa$--gauge fixing. Due to (7), in fact, the matrix
$\Gamma$ defined in (4) satisfies $\Gamma^2=\unity$ and $tr\ \Gamma=0$, 
such that the constraint
$$
{\unity + \Gamma\over 2} y=0\eqno(8)
$$
eliminates just half of the sixteen $y's$ and defines an $SO(1,9)$ 
covariant background gauge fixing$^{[20]}$. Moreover, (7) allows 
also to write an embedding equation for the $SO(1,5)$ (torsion free) spin 
connection one--form $\omega_{\widehat a\ \widehat b}=d\sigma^j 
\omega_{j\widehat a\ \widehat b}$
$$
\omega_{j\widehat a\ \widehat b} = \omega^{(0)}_{j\widehat a \ \widehat b} 
+e^{{1\over 3}\varphi}
\left( {1\over 2} v^B_{\widehat b} v^C_{\widehat a} v_{ja} + 
v_{[\widehat b}{}^B 
v_{\widehat a]a} v^C_j\right)
T_{CB}{}^a + {2\over 3} e_{j[\widehat a} e^i_{\widehat b]}
\partial_i\varphi\eqno(9)
$$
where
$$\eqalign{
\omega^{(0)}_{j\widehat a\ \widehat b} & = \left( \partial_j v^a_{[\widehat a}
-\Omega_j{}^{ab} v_{b[\widehat a}\right) v_{\widehat b]a}\cr
v^A_{\widehat a} & = e_{\widehat a}^j \ v_j^A\cr
\Omega_{jab} & = V^A_j \Omega_{Aab}.\cr}
$$
Notice that $\omega-\omega^{(0)}$ is an $SO(1,5)$ {\it tensor} one--form,
meaning that 
the anomaly polynomials associated to $\omega$ and $\omega^{(0)}$ fall 
in the same $SO(1,5)$--anomaly cohomology class.
We will take advantage of this fact below.
The $SO(1,5)$ and $SO(1,9)$ curvature two--forms are given respectively by
$$
\eqalign
{
{\cal R}_{\widehat a}{}^{\widehat b} &= d\omega_{\widehat a}{}^{\widehat b} +  
\omega_{\widehat a}{}^{\widehat c}\omega_{\widehat c}{}^{\widehat b}
\cr
\R_a{}^b&=d\Omega_a{}^b+\Omega_a{}^c\Omega_c{}^b.
\cr
}
$$
To be precise, the former is an intrinsic $d=6$ form while the latter,
as it stands, is a superform of the ten dimensional target superspace, 
whose pullback on the $d=6$ worldvolume is  naturally induced
by $dZ^M=d\sigma^j\partial_jZ^M$. In what follows we will not indicate
this pullback explicitly since its occurrence will be clear from the
context.

\vskip 0.5truecm

\noindent
{\bf 3. The target space Lorentz anomaly}

\vskip 0.5truecm

Upon  performing the normal coordinate expansion$^{[19]}$ of the action (1) 
and using (5) one gets for the $SO(1,5)$ and $SO(1,9)$ invariant kinetic 
term of the $y^\alpha$ the expression
$$
{1\over 2} \int d^6\sigma\sqrt{g} \ e^{-{1\over 3}\varphi}
e^j_{\widehat a}\  v_a^{\widehat a}\ y\ \Gamma^a {\unity-\Gamma\over 2} D_j y
$$
\noindent
where $D_j = \partial_j - {1\over 4} \Gamma_{cd} \Omega_j{}^{cd}$. 
Enforcing the gauge fixing and rescaling the $y's$ this becomes
$$
I(v,\Omega,y) = {1\over 2} \int d^6\sigma \sqrt{g} \ e^j_{\widehat a} \
v^{\widehat a}_a\ y\ \Gamma_a {\unity-\Gamma\over 2} D_j 
{\unity-\Gamma\over 2} y.\eqno(10)
$$
Actually, the normal coordinate expansion of the action (1) gives rise
to additional terms quadratic in the $y^\alpha$'s which in eq. (10) we did
not write. 
The only effect of these additional terms is a redefinition of the 
$SO(1,9)$ Lorentz connection by 
Lorentz {\it covariant} terms. Therefore they give rise at most to trivial 
anomalies (trivial cocycles) and can be neglected.
Eq. (10) is the starting point of our perturbative analysis of
the super--fivebrane anomalies.
As for the string, the non--canonical dependence of the 
propagator on the $v_a{}^{\widehat a}$ can be eliminated by an $SO(1,9)$ 
rotation of the $y^\alpha$ which is, however, expected to be anomalous. So 
the first step consists in deriving the $SO(1,9)$ anomaly associated to 
(10). Since the form of the anomaly is strongly constrained by the 
consistency condition  it is sufficient to determine the anomaly under an 
$SO(4)$ subgroup of $SO(1,9)$, and for a particular configuration of the 
background fields. We choose a configuration for which
$$
\eqalignno
{
e_{\widehat a}{}^j &= \delta_{\widehat a}{}^j & (11)\cr
v^a_{\widehat a} (\sigma) &= {\rm constant}&(12)\cr
v_{\widehat a}^a \ \Omega_{jab} &=0.&(13)\cr
}
$$
This implies in particular that the $16\times 16$ matrices $\Gamma$ and 
$\Gamma^j \equiv e^j_{\widehat a} \ v_a^{\widehat a} \ \Gamma^a=
g^{ji}v_{i a}\Gamma^a$,
appearing in 
(10), are now {\it constant} matrices  and satisfy a six--dimensional Dirac 
algebra:
$$\eqalign{
\{ \Gamma^i, \Gamma^j\} = 2\eta^{ij}\cr
\{ \Gamma, \Gamma^i\} = 0.\cr}\eqno(14)
$$
Moreover, due to (13)
$$
\Omega_{jab} \cdot [\Gamma^{ab}, \Gamma^i] = 0.\eqno(15)
$$
For this particular configuration the action (10) is invariant under a local 
$SO(4)$ subgroup of $SO(1,9)$ specified by the constraint
$$
v_{\widehat a}^a \ L_{ab}=0.\eqno(16)
$$
For the configuration (11)--(13) the $y$--propagator becomes just
${k^j\Gamma_j\over k^2}$ and the $y$-$y$-$\Omega$ vertex is 
$i{1-\Gamma\over 2} \left({1\over 4} \Gamma_{ab}\right)$. 
Then the leading one--loop 
anomaly diagrams, which in six dimensions are box diagrams with four 
external $\Omega$, can be easily evaluated. Calling 
$\R^{(0)}_{ab}=d\Omega_{ab}$, 
the anomaly coming from the box diagrams can be 
computed to be
$$
{1\over 384} {1\over (2\pi)^3} \int \left( tr \left(L\R^{(0)} \R^{(0)} 
\R^{(0)}\right)
- {3\over 4} tr \left(L\R^{(0)}\right) tr\left(\R^{(0)} \R^{(0)}\right)\right),
\eqno(17)
$$
where the traces are in the fundamental representation of $SO(1,9)$. To get
this one has to use that in the Feynman diagrams, thanks to (14,15), one
can replace the $\Gamma$--matrix traces
$$
tr \Bigl[ (\Gamma^i \Gamma^j\cdots) (\Gamma^{ab}\Gamma^{cd}\cdots)\Bigr] 
\rightarrow {1\over 16}\ tr (\Gamma^i\Gamma^j\cdots)\ tr (\Gamma^{ab}
\Gamma^{cd}\cdots)\eqno(18a)
$$
and that
$$
tr \left(\widetilde H \widetilde H \widetilde H \widetilde H\right) = 
-  tr\left(H^4\right) +
{3\over 4} \left(tr\ H^2\right)^2\eqno(18b)
$$
for any antisymmetric matrix $H^{ab}$ where $\widetilde H \equiv {1\over 4} 
\Gamma_{ab} H^{ab}$, and the traces at the r.h.s. of (18b) are in the 
fundamental representation of $SO(1,9)$. From (17) one can read the 
anomaly polynomial associated to $SO(1,9)$ transformations as 
$$
X^L_8 = {1\over 384} {1\over (2\pi)^3}
\left( tr\R^4- {3\over 4} \left(tr \R^2\right)^2 + \gamma \ tr \R^2  tr
{\cal R}^2\right).\eqno(19)
$$
We included a term proportional to $tr {\cal R}^2$
which could not be derived by the method 
above since in the configuration (11)--(13) 
${\cal R}_{\widehat a\ \widehat b}=0$. 
The unknown coefficient $\gamma$ will be determined below.

The result (19), for $\gamma=0$, can actually also be derived  using the
index theorem$^{[23]}$. For a complex Weyl fermion in $d=6$ the 
anomaly polynomial is given by 
$$
X^I_8 = {1\over 384} {1\over (2\pi)^3}
\left( -16\ tr \F^4+4\ tr \F^2  tr{\cal R}^2 -
{N\over 12} \left(tr {\cal R}^2\right)^2-{N\over 15} \ tr {\cal R}^4\right),
\eqno(20)
$$ 
where the Yang--Mills trace over the $\F'$s is in whatever representation the
fermions are and $N$ is its dimensionality. In the configuration (11)--(13)
the action (10) corresponds indeed to one real chirally  projected fermion 
with sixteen components, $y^\alpha$, which is equivalent to one complex 
$d=6$ Weyl fermion. The ``Yang--Mills" matrices $\F$
are replaced by ${1\over 4}
\Gamma_{ab}\R^{ab}$, but, since these matrices and the $d=6$ Dirac matrices 
$\Gamma^i$ live in the same 16--dimensional representation space, the traces
over the two kinds of matrices factorize with a factor of ${1\over 16}$ (see
(18a)). In summary, one has to set in (20)
$$\eqalign{
N & =1\cr
{\cal R} & = 0 \cr
tr\ \F^4 & \rightarrow {1\over 16} \ tr
\left( {1\over 4} \Gamma_{ab} \R^{ab}\right)^4,\cr} \eqno(21)
$$
and, due to (18b), one gets again (19) with $\gamma=0$.

\vskip 0.5truecm

\noindent
{\bf 4. The worldvolume Lorentz anomaly}

\vskip 0.5truecm

Eq. (19) parametrizes the non--invariance 
of the measure $\int \{ Dy\}$ in the functional 
integral which defines the effective action, under a
generic finite $SO(1,9)$ rotation on the $y^\alpha$, with
transformation matrix $\Lambda^a{}_b \in SO(1,9)$. Performing such 
a rotation we can rewrite the effective action $\Gamma$ as
$$
\eqalign
{
e^{i\Gamma} & = \int \{ Dy\} e^{iI(v, \Omega,y)}\cr
& = e^{-i\Gamma_{WZ}(\Lambda)} \int \{ Dy\} 
\ e^{iI(v^\Lambda, \Omega^\Lambda, y)}\cr 
& \equiv e^{-i\Gamma_{WZ}(\Lambda)} \ e^ { i \Gamma_0},
\cr
}
\eqno(22)
$$
where
$$
\eqalign
{
\Omega^\Lambda&=d\Lambda\Lambda^T + \Lambda\Omega\Lambda^T\cr
\left(v^\Lambda\right)^a_{\widehat a}&=\Lambda^a{}_b\ v^b_{\widehat a}\ ,\cr
}
$$
and the Wess--Zumino term $\Gamma_{WZ}(\Lambda)$ 
can be read off from  (19) as follows. If we 
define for a generic $SO(M)$--connection one--form 
$\widetilde\Omega$
and the related curvature two--form, ${\widetilde {\R}} 
= d{\widetilde{\Omega}} + {\widetilde{\Omega}}{ \widetilde{\Omega}} $,
canonical Chern--Simons forms,
$$
\eqalign
{
tr{\widetilde {\R}}^2 &= dY_3 ({\widetilde{\Omega}})\cr
tr{\widetilde {\R}}^4 &= dY_7 ({\widetilde{\Omega}}),\cr
}\eqno(23)
$$
then, due to (19) and (22), we have 
$$
\Gamma=\Gamma_0 - \int_{M^7} \left(U_7\left(\Omega^\Lambda\right) 
- U_7(\Omega)\right),
\eqno(24)
$$
where
$$
U_7(\Omega) = {1\over 384(2\pi)^3} \left(Y_7(\Omega) 
- {3\over 4} Y_3 (\Omega) \ 
tr \R^2 + \gamma\ Y_3 (\Omega)\ tr {\cal R}^2\right). \eqno(25)
$$
The boundary of $M_7$ is the fivebrane worldvolume.
Until now we considered a generic $\Lambda \in SO(1,9)$. We want now choose
a $\Lambda$ for which the kinetic term of the $y's$ in $I (v^\Lambda, 
\Omega^\Lambda, y)$ becomes  canonical, allowing to read the 
worldvolume Lorentz anomaly of 
$\Gamma_0$ directly from the index theorem. To do this we choose a basis in
the four--dimensional space orthogonal to the six $v_{\widehat a}{}^a$, 
introducing four $SO(1,9)$ vectors $\{N^r_a\}, r,s = 6,7,8,9$, satisfying
$$\eqalign{
N_a{}^r N^{as} & = -\delta^{rs}\cr
N_a{}^r v_{\widehat a}^a  & = 0,\cr}\eqno(26)
$$
and set $\left(\Lambda^b{}_a\equiv\{\Lambda^{\widehat b}{}_a, 
\Lambda^r{}_a\}\right)$
$$\eqalign{
\Lambda^{\widehat b}{}_a & = v^{\widehat b}_a\cr
\Lambda^r{}_a & = N_a^r.\cr}\eqno(27)
$$
This $\Lambda$ belongs indeed to $SO(1,9)$ in that $\Lambda^a{}_b \Lambda^c
{}_d \eta^{bd} = \eta^{ac}$, due to (7) and (26).

This procedure introduces an ``intermediate" external local symmetry 
group $SO_E(4)$
in the game, with $SO(1,5)$ and $SO(1,9)$--invariant connection one--form 
given by
$$
d\sigma^j W_{jrs} = W_{rs} = \left(dN_{ra}-\Omega_{ab} N^b_r\right) N_s^a,
$$
whose curvature is
$$
T_{rs} = d W_{rs} + W_r{}^t W_{ts}.
$$
We proceed now to the determination of the $SO(1,5), SO_E(4)$ and
$SO(1,9)$ anomalies coming from each of the three terms in (24). 
$U_7(\Omega)$ carries only $SO(1,9)$ anomalies with anomaly polynomial
given clearly by $X_8^L$,
see (19). $\Gamma_0$ and $\int U_7\left(\Omega^\Lambda\right)$, on the 
contrary, carry only $SO(1,5)$ and $SO_E(4)$ anomalies and, moreover, since
$SO_E(4)$ is an intermediate symmetry group $\Gamma_0-\int U_7 
(\Omega^\Lambda)$ has to be $SO_E(4)$ invariant (this last condition will 
allow us eventually to determine the coefficient $\gamma$ in eq. (19)).
Let us first give the results. The anomaly polynomial carried by 
$-\int U_7\left(\Omega^\Lambda\right)$ turns out to be
$$
X^\Lambda_8 = {1\over 384 (2\pi)^3} 
\left(-tr {\cal R}^4 - tr T^4 + \left( {3\over 4} - \gamma\right) 
\left(tr {\cal R}^2\right)^2 + {3\over 4} \left(tr T^2\right)^2 
+ \left( {3\over 2}-\gamma\right)
tr {\cal R}^2 tr T^2\right), \eqno(28)
$$
while that associated to $\Gamma_0$ is
$$
X^{(0)}_8 = {1\over 384(2\pi)^3} \left( - {1\over 15} tr {\cal R}^4 + 
tr T^4 - {1\over 12} \left(tr {\cal R}^2\right)^2 - {3\over 4} 
\left(tr T^2\right)^2 +
{1\over 2} tr {\cal R}^2 tr T^2\right), \eqno(29)
$$
where all traces are in the fundamental representations of the 
$SO$--Lie algebras.
To get (28) one has to note that
$$
\eqalign{
(\Omega^\Lambda)_{\widehat a \ \widehat b} & = \omega^{(0)}_{\widehat a
\ \widehat b}\cr
(\Omega^\Lambda)_{rs} & = W_{rs}, \cr}\eqno(30)
$$
while $(\Omega^\Lambda)_{\widehat a r} = (d v_{\widehat a}^a - \Omega^{ab}
v_{b\widehat a}) N_{ar}$ is not a connection but a {\it tensor} 
under $SO(1,5)\otimes
SO_E(4)$, and to use the decompositions
$$
\eqalign{
Y_7(\Omega^\Lambda) & = Y_7 \left(\omega^{(0)}\right) + Y_7 (W) + X_7 + dX_6\cr
Y_3(\Omega^\Lambda) & = Y_3 \left(\omega^{(0)}\right) + Y_3(W) + X_3\cr
tr\R^2 & = tr {\cal R}^2 + trT^2 + dX_3.\cr}\eqno(31)
$$
Here $X_3$ and $X_7$ are  {\it completely invariant} forms, given in 
the appendix, and $X_6$ is a {\it local} form which can, therefore, be
disregarded. Taking (31) into account it is straightforward to compute 
$\delta\left(-\int U_7\left(\Omega^\Lambda\right)\right)$ and 
to realize that the resulting 
anomaly, apart from trivial cocycles, is represented by $X^\Lambda_8$.
To compute $X^{(0)}_8$ we observe that, with the choice (27), one has
$$
I(v^\Lambda, \Omega^\Lambda, y) = {1\over 2} \int d^6\sigma
\sqrt{g} \ y \ e^j_{\widehat a}\ \Gamma^{\widehat a}\ 
{\unity-\widehat\Gamma\over 2}
\left( \partial_j - {1\over 4} \omega^{(0)}_{j\widehat b \ \widehat c}
\ \Gamma^{\widehat b\ \widehat c} - {1\over 4} W_{jrs} \ \Gamma^{rs}\right)y,
\eqno(32)
$$
where now all gamma matrices are constant matrices and 
$\widehat\Gamma=\Gamma^0\Gamma^1 \cdots\Gamma^5$ gives rise to a true chiral 
projector, ${1-\widehat\Gamma\over 2}$. The kinetic term for the $y'$s 
in (32) 
is now canonical and the anomalies carried by $\Gamma_0$ can be retrieved
from 
the index theorem, eq. (20). Our fermion  is sixteen--dimensional, but 
real, corresponding thus to {\it one} complex $d=6$ Weyl fermion,  
hence we have to set $N=1$.
Our gauge--group is $SO_E(4)$, with generators ${1\over 4} \Gamma^{rs}$, 
hence we have to identify $\F\rightarrow {1\over 4}\Gamma^{rs} T_{rs}$; but
since the matrices $\Gamma^{\widehat a}$ and $\Gamma^{rs}$, although being 
commuting, live in the same sixteen--dimensional space, the traces over the 
$\F'$s
have to be divided by 16. Taking the identity (18b) into account, (with
$\Gamma_{ab}\rightarrow\Gamma_{rs}$), one gets indeed (29). 

In summing up $X^L_8, X^\Lambda_8$ and $X_8^{(0)}$ one sees that the
cancellation of the 
$SO_E(4)$ anomalies requires  
$$
\gamma=2.
$$ 
(This cancellation could, 
actually, also be used as an alternative procedure to 
determine the $SO(1,9)$ anomaly
polynomial $X^L_8$ completely).
The total anomaly polynomial arising from  the gravitational sector
of the super--fivebrane sigma--model is thus 
given by
$$\eqalign{
X_8 & = X^L_8 + X^\Lambda_8 + X^{(0)}_8\cr
&= {1\over 384(2\pi)^3}  \left( - {16\over 15} \ tr {\cal R}^4 - {4\over 3}
\left(tr {\cal R}^2\right)^2 + 2\ tr {\cal R}^2 tr \R^2 - {3\over 4} 
\left(tr \R^2\right)^2 + 
tr \R^4\right).\cr} \eqno(33)
$$

\vskip 0.5truecm

\noindent
{\bf 5. Adding the anomaly from the ``heterotic" sector}

\vskip 0.5truecm

Eq. (33) is our principal result. We see that even for a flat $D=10$ 
background, i.e. for $\R=0$, the worldvolume $SO(1,5)$ anomaly is non 
vanishing. To cancel this anomaly one has necessarily to add a ``heterotic"
sector to the theory. As said in the introduction, it is still unknown how 
to couple such a sector in a $\kappa$--invariant way. If we assume, in 
analogy to the string, that such a sector is made out of $N_\psi$ $d=6$ complex 
Weyl fermions, with chirality opposite to that of the $y^\alpha$, which 
belong to an $N_\psi$--dimensional representation of a gauge group $G$, and 
that they are chirally coupled to the gauge fields ${\rm A}$, with Lie algebra
valued curvature two--form $\F=d{\rm A}+{\rm A}{\rm A}$,
 and to the worldvolume 
connection $\omega$, then one has to add to $X_8$ just  $-X^I_8$ 
with $N=N_\psi$ (see eq. 
(20)). The resulting anomaly polynomial of the ``heterotic" fivebrane 
would then be given by $X^H_8 = X_8 - X^I_8$ which can be written as
$$
\eqalign
{
X^H_8 = {1\over (2\pi)^3} {1\over 384}& \left( (N_\psi-16) \left( {1\over 15}tr
{\cal R}^4 +{1\over 12} \left(tr {\cal R}^2\right)^2\right)\right.\cr
&+ \left(2 tr {\cal R}^2 -
tr \R^2\right) \left(tr \R^2 - 2 tr\F^2\right)
\cr
& \left.  + tr \R^4 + {1\over 4} 
\left(tr \R^2\right)^2 - 2 tr \F^2 tr \R^2 + 16 tr
\F^4\right). 
\cr
}
\eqno(34)
$$
First of all we observe that for a trivial background $(\R=\F=0)$ the
cancellation of the $SO(1,5)$ anomalies requires sixteen heterotic 
fermions,
$$
N_\psi=16. \eqno(35)
$$
For a non--trivial background the anomalies which do not involve
${\cal R}$, but only the target space curvatures $\R$ and $\F$, can be cancelled
by choosing an appropriate transformation law for $B_6$ and modifying
accordingly its field strength by an appropriate Chern--Simons form,
(see ref. [13]). The resulting $D=10$ supergravity theory would contain
the Yang-Mills field strength $\F$ with values in a 
{\it sixteen--dimensional} 
representation of some gauge group $G$: this fact prevents already 
a direct comparison with heterotic string theory since neither 
$SO(32)$ nor $E_8\otimes E_8$ admit a sixteen-dimensional irreducible
representation.

The result (34) presents also a feature which is more dramatic for
the fivebrane itself: 
the mixed term in the second line,
proportional to $tr {\cal R}^2(tr \R^2-2tr \F^2)$,
can not be cancelled by modifying the $B_6$-field strength,
and the heterotic fivebrane sigma--model would be inconsistent at the
quantum level. 

To discuss our results more in detail we
recall now the principal features of string/fivebrane duality.

\vskip 0.5truecm

\noindent
{\bf 6. The string/fivebrane duality conjecture}

\vskip 0.5truecm

The pure $N=1,D=10$ supergravity theory in ten dimensions admits 
two dual formulations, one based on $B_2$ and one based on $B_6$, 
whose dynamics is described by symmetric formulations in that
both field strengths are closed, $dH_3=0=dH_7$. This symmetry
disappears if one couples the pure Supergravity to a 
Super--Yang--Mills theory because in this case
$dH_3\neq 0$; in some sense it has been restored 
through the discovery of superstring theories, more precisely through the
Green--Schwarz anomaly cancellation mechanism which implies $dH_3\neq 0,
dH_7\neq 0$. 
The $N=1,D=10$ Supergravity--Super--Yang--Mills anomaly
polynomial $I_{12}$ factorizes, in fact, 
 for the gauge 
groups $SO(32)$ and $E_8\otimes E_8$ into $I_{12}= {1\over 2\pi} I_4 I_8$
$^{[3]}$ where for $SO(32)$ one has
$$
\eqalignno
{
I_4
& = {1\over (2\pi)}  {1\over 4} \left(tr \R^2 - tr{\cal F}^2\right)
\equiv d\omega_3
&(36)\cr
I_8 &
= {1\over (2\pi)^3} {1\over 192} \left(tr\R^4 + {1\over 4}
(tr \R^2)^2 - tr \R^2 tr{\cal F}^2 + 8\ tr {\cal F}^4\right)\equiv 
d\omega_7, 
&(37)\cr
}
$$
where the traces over the ${\cal F}'$s are in the fundamental representation of 
$SO(32)$. On the other hand,
the cancellation of  sigma--model $\kappa$--anomalies
in the heterotic string, with action normalized in a standard way as
$S_2 = - {1\over 2\pi\alpha^\prime}$ $\int d^2\sigma {1\over 2} \varepsilon^{ij}
V_i^A V^C_j B_{CA}+\cdots$, requires the introduction of an invariant 
$B_2$--field strength given by$^{[19]}$ 
$$
\eqalign
{
H_3 &= dB_2 - (2\pi\alpha^\prime)\omega_3\cr
dH_3& = - (2\pi\alpha^\prime) I_4.\cr
}
\eqno(38)
$$
As a consequence $B_2$ transforms anomalously under $SO(1,9)$ and
$SO(32)$ and the $N=1,D=10$ 
 anomaly can be cancelled by adding to the classical supergravity action 
the term
$$
\Delta S_{10} = - {1\over 2\pi} \int
\left( {1\over 2\pi\alpha^\prime} B_2 I_8 + {2\over 3} \omega_3 
\omega_7\right).\eqno(39)
$$
Since the kinetic term  of $B_2$ is given
by  $S_{10} = - {1\over 2\kappa^2} \int {1\over 2} e^{-2\varphi} H_3 * 
H_3$, where $\kappa^2$ is the ten dimensional Newton's constant, the 
addition of 
(39) modifies the field equation of $H_3$ to
$$
d\left(* e^{-2\varphi} H_3\right) = {2\kappa^2\over (2\pi)^2 \alpha^\prime} 
I_8,
\eqno(40)
$$
which represents a one--loop string effect.
According to the string--fivebrane duality conjecture, eq. (40)
should arise from the cancellation of sigma--model anomalies in the 
heterotic fivebrane, as the Bianchi identity for the generalized $B_6$
field--strength, via the identification
$$
H_7 = * e^{-2\varphi} H_3,
$$
which leads to
$$
\eqalignno
{
dH_7& = {2\kappa^2\over (2\pi)^2\alpha^\prime} I_8 &(41)\cr
H_7 &= dB_6 + {2\kappa^2\over (2\pi)^2\alpha^\prime} \omega_7, &(42)\cr
}
$$
where $B_6$ is identified with our fivebrane six--form in (1).

Another characteristic feature of this duality is the 
Dirac--Nepomechie--Teitelboim quantization condition$^{[24]}$ 
on the tensions of strings and fivebranes 
which reads
$$
2\kappa^2 = n(2\pi)^5\alpha^\prime\beta^\prime, \eqno(43)
$$
where $n$ is integer.

\vskip 0.5truecm

\noindent
{\bf 7. Discussion}

\vskip 0.5truecm

The duality conjecture leads us to compare 
$I_8$ in eq. (37) with the target space anomaly polynomial arising
from eq. (34). If we set $N_\psi=16$ and disregard for the moment the second
line of (34) we are led to compare its third line with $I_8$. As we
mentioned already, this comparison is prevented by the fact that
the representations of $\F$ and ${\cal F}$ do not match, but apart
from that we see also that the terms 
not involving $\F$ carry coefficients which differ
by a factor
of 1/2 w.r.t. the corresponding terms in $I_8$. 
So one is forced to conclude that a
fundamental super--fivebrane described by the action (1), supplemented with a
heterotic sector of 16 complex Weyl fermions, with still unknown action, is not
in agreement with string--fivebrane duality. We can, however, make the 
following observations.

\vskip 0.3truecm\noindent
{\it Doubling the gravitational fivebrane sector?}
It is worthwhile to notice that, for the gauge group $SO(32)$,
we could find complete agreement 
  with the duality conjecture if the gravitational sector of the 
fivebrane would correspond to {\it two}, instead of one, complex Weyl 
fermions. 
For what concerns the anomaly this would just amount to 
multiply the anomaly polynomial $X_8$ in (33) by a factor of two,
and, upon adding the heterotic sector, one would obtain for the
total anomaly polynomial, instead of (34),
$$
\eqalign
{
{\widetilde X}^H_8 =2X_8-X_8^I 
= {1\over (2\pi)^3}{1\over 192} &\left( (N_\psi-32) 
\left( {1\over 15} tr
{\cal R}^4 +{1\over 12} \left(tr {\cal R}^2\right)^2\right)\right.\cr
&+ (2 \ tr {\cal R}^2 -
tr \R^2) (tr \R^2 -  tr \F^2)
\cr
& \left.  + tr\R^4 + {1\over 4} 
\left(tr \R^2\right)^2 -  tr \F^2 tr \R^2 + 8 \ tr 
\F^4\right). 
\cr
}
\eqno(44)
$$
In this case on needs indeed 32 heterotic fermions to cancel the
pure worldvolume anomaly, the fivebrane gauge group can be taken to be
$SO(32)$ and one can identify $\F\equiv {\cal F}$. Moreover, the second
line in (44) can now be eliminated by invoking the anomalous
transformation law of $B_2$ resulting from (38) and 
adding to the classical fivebrane action the local term
$$
\eqalign
{
\Delta S_6 &= {1\over (2\pi)^3}{1\over 192}  {4\over \alpha^\prime}
\int B_2 (2\ tr\ {\cal R}^2 - tr\ \R^2)\cr
 &= {1\over (2\pi)^3}{1\over 192} 
{4\over \alpha^\prime} \int (H_3+ 2\pi\alpha^\prime\omega_3) 
(2Y_3(\omega)-Y_3(\Omega)).\cr
}
\eqno(45)
$$
The advantage of the second form of this counterterm is that it does not 
involve
the ``string" two--form $B_2$, but only the curvature
$H_3= *e^{2\varphi}H_7$. Its variation cancels then the second
(dangerous) line in (44) upon using the $H_3$--Bianchi identity in (38)
which has to be interpreted as equation of motion for $H_7$. What 
remains of ${\widetilde X}_8^H$ is then just a pure target space polynomial,  
the third line in (44), which coincides exactly with $I_8$. The 
corresponding anomaly can then be eliminated by setting
$$
\eqalignno
{
H_7 &=dB_6 + (2\pi)^3\beta^\prime \omega_7 &(46)\cr
dH_7 & = (2\pi)^3\beta^\prime I_8.& (47)\cr
}
$$
These equations would then perfectly respect the duality conjecture
since they coincide with (42,41) if (43) holds with $n=1$. 
Notice, however, that as a consequence of the necessary subtraction 
of the local term (45),
and contrary to what happens for the string sigma--model, here the anomaly
cancellation mechanism
requires not only the Bianchi identity for $H_7$ but also its 
equation of motion (38). While the solutions of the
Bianchi identity  (38) in superspace are well known
until now no solution is known for (47); apart from that one should also
keep in mind that if one insists on both, regarding one as equation of
motion and the other as Bianchi identity, even if one can solve them 
simultaneously in superspace, the resulting supergravity
 equations of motion can not
be deduced from a local gauge-- and Lorentz--invariant supergravity action. 

\par
It is certainly difficult to imagine that a consistent fivebrane 
sigma--model exists in which the gravitational anomaly (33) is just
doubled; nevertheless our quantitative result -- i.e. that the
anomaly coming from the gravitational sector of the fivebrane 
is just half of what would be naively expected on the basis of 
the duality conjecture -- for which at present we have no 
clear interpretation, may in the future help to cast the
duality conjecture itself in a more concrete formulation.
\vskip 0.3truecm

\vskip 0.3truecm\noindent
{\it Particular configurations of the gauge fields.} It may be interesting
to notice that if one insists on 16 heterotic fermions, and hence on
(34), from a purely formal point of view one can find agreement with
duality if one chooses particular configurations for the Yang--Mills
fields. First of all, if on sets them to zero ($\F= {\cal F}=0$),
the second line in (34) can be eliminated by subtracting now
from the classical action $1/2\Delta S_6$ and imposing
$dH_7={\beta^\prime\over 384}
\left(tr \R^4 + {1\over 4} \left(tr \R^2\right)^2\right)$. This matches
now with (41), for ${\cal F} =0$, if 
$$
2\kappa^2 = 1/2(2\pi)^5\alpha^\prime\beta^\prime \eqno(48)
$$ 
which corresponds to (43), but with $n=1/2$. Since the Dirac quantization
condition arises from the requirement that the product of the
charges of a single fivebrane and a single string are integer, (48)
would amount 
to the existence of half--charged elementary fivebranes. 
Half--charged fivebranes arose, actually, in ref. [25]
where they appear, however, always in pairs such that their total
charge is always integer. Half integral magnetic charges have arisen
also on fixed points of $Z_2$-orbifold compactifications of $N=1,D=11$
Supergravity in ref. [26].

Incidentally one may notice that one can cancel (34) also if
one couples the 16 heterotic fermions to 16 abelian gauge fields and 
introduces an $H_3$ satisfying 
$$
dH_3=-{\alpha^\prime\over 4}\left(tr\R^2-2tr\F^2\right).
\eqno(49)
$$
$X_8^H$ could then be cancelled subtracting 
$1/2\Delta S_6$ and setting $dH_7$ equal to the 
third  line of (34). It is puzzling to notice that the resulting 
equation for $H_7$ and (49) coincide with
(41) and (38) respectively, imposing again (47),
if one sets in ${\cal F}$ the 480 non abelian 
gauge fields of $SO(32)$ to zero and identifies the ones in its
Cartan subalgebra with the 16 abelian gauge fields to which the
heterotic fermions are coupled.

\vskip 0.3truecm\noindent
{\it Comment on the target space polynomial of ref. [13]}
The results of the present paper, and of ref. [20], allow us to give
a partial justification of the somehow conjectural
derivation of the {\it target space} anomaly polynomials for the heterotic
string and fivebrane sigma--models performed by Dixon et. al. in [13].
The method implied in that paper works, in fact, once one has made
sure of the cancellation of {\it worldsheet/worldvolume} anomalies.
For the string we showed in [20] that the worldsheet Lorentz anomalies 
get a contribution $N_y=8$, from the eight physical Majorana--Weyl
quantum $\vartheta$'s, and a contribution $L=24$
induced by the $D=10$ target space Lorentz anomaly, which has weight
$-L$, via a 
Wess--Zumino term. This anomaly is cancelled introducing $N_\psi=32$
heterotic fermions such that
$$
L=N_\psi-N_y. \eqno(50)
$$
This equation implies that one can compute the target space Lorentz
anomaly by using formally the index theorem and counting the
fermions, taking their chirality into account, as $L=32-8$. This
was, indeed, the procedure applied in [13] and the result was 
$I_4$ of eq. (36). The reason for why this works for the string 
exactly is that in two dimensions 
the anomaly polynomial contains only irreducible invariants i.e. $tr\F^2$
and $tr \R^2$.

For the fivebrane the same reasoning can be applied 
for what concerns the irreducible term $tr\R^4$. We saw that in this
case one has $N_y=1$ and $L=15$, see eqs. (29,28), and the cancellation
of the worldvolume anomalies required $N_\psi=16$. This means that
the weight of $tr\R^4$ can be computed formally using the index theorem (20), 
with $tr{\cal R}^4\rightarrow tr\R^4$ and $N=-L=1-16$, in agreement
with (19). In ref. [13] an overcounting of the quantum $\vartheta$'s
led to $N_y=2$, and the heterotic fermions have been assumed to be 32 
instead of 16; this led to $-L=2-32$ and brought to a doubling
of the coefficient of $tr\R^4$. For what concerns the factorized terms
in the anomaly polynomial we can only observe that the polynomial
(44), in which the $y'$s have been doubled by hand 
and  one should set $N_\psi=32$, 
reduces to the corresponding expression in [13] only if,
instead of setting ${\cal R}=0$ as one should, one 
identifies $tr{\cal R}^2 \leftrightarrow tr\R^2$. 

\vskip 0.3truecm

In conclusion, our results add to the well known open problem of a
$\kappa$--invariant action for the heterotic
sector of the fivebrane a further one, namely, that the 
worldvolume and target space anomalies of its gravitational
sector are 1/2 of what would be expected on the basis of
string/fivebrane duality. Sometimes two difficulties which at first sight 
seem unrelated conspire to shed new light on both of them. Our hope
is that the analysis presented here will contribute in the future
to find a consistent formulation for a heterotic super--fivebrane
sigma--model.

\vskip 0.5truecm
\noindent
{\bf Appendix}

\vskip 0.5truecm

\noindent
I) The symmetric tensor $X^{ij}$, which parametrizes the 
$\kappa$--transformation of the metric, is given by 

$$\eqalign{
X^{ij} & = - 2 Y^{ij} (v^{ka} v^\beta_k(\Gamma_a)_{\alpha\beta}
+\lambda_\alpha)\kappa^\alpha - {1\over 3} g^{ij} \lambda_\alpha 
\Delta^\alpha\cr
+ {2\over 5!\sqrt{g}} & \varepsilon^{j_i\cdots j_5(j}v^{i)\alpha}
v^{a_1}_{j_1}\cdots v^{a_5}_{j_5}
(\Gamma_{a_1\cdots a_5})_{\alpha\beta}\Delta^\beta\ ,\cr}
$$
where
$$
Y^{ij} \equiv {1\over 6!}
{\varepsilon^{ij_1...j_5}\over\sqrt{g}} {\varepsilon^{jj_1...j_5}\over\sqrt{g}}
 (Z_{i_1j_1}\cdots Z_{i_5j_5} + Z_{i_1j_1}\cdots Z_{i_4 j_4} g_{i_5 j_5}+
\cdots + g_{i_1 j_1} \cdots g_{i_5 j_5})
$$
and
$$
Z_{ij} = v_i^a v_{ja}.
$$

\smallskip
\noindent
II) The completely invariant forms $X_3$ and $X_7$ can be expressed in 
terms of the following $SO_E(4)\otimes SO(1,5)$ tensor--forms:

$$\eqalign{
C_{\widehat a r} & = (\Omega^\Lambda)_{\widehat a r}\cr
(C^T)_{r\widehat a} & = - C_{\widehat a r}\cr
P_{\widehat a r} & = dC_{\widehat a r} + W_r{}^s C_{\widehat a s} + 
\omega_{\widehat a} 
{}^{\widehat b} C_{\widehat b r} \cr
(P^T)_{r\widehat a} & = - P_{\widehat a r}. \cr}
$$
We have:

$$\eqalign{
X_3 =& 2  tr \left(CP^T\right)\cr
X_7 =& tr \Bigl(4 \left(C^TC\right)^2 C^T P+2 PP^T PC^T + 
2 C^T CP^T CT\cr
& + 2CC^T PC^T {\cal R} + 4 C^T CP^T {\cal R} C+4 CC^T PTC^T\cr
& + 4 {\cal R} PTC^T + 4 T^2 P^TC + 4{\cal R}^2 PC^T\Bigr),\cr}
$$
where the traces here mean simply contractions of the indices.

\vskip 0.5truecm

\noindent
{\bf References}

\vskip 0.5truecm

\item{[1]} A. Chamseddine, Nucl. Phys. {\bf B185}, 403 (1981).

\item{}G.E. Chaplin and N.S. Manton, Phys. Lett. {\bf 120B}, 105 (1983).

\item{}B.E.W. Nilsson, Nucl. Phys. {\bf B188}, 176 (1981); Goteborg preprint 81--6
(1981).

\item{}R. D'Auria, P. Fr\`e and A.J. da Silva, Nucl. Phys. {\bf B196}, 205 (1982).

\item{}E. Bergshoeff, M. de Roo, B. de Wit and P. van Nieuwenhuizen, Nucl. Phys. 
{\bf B195}, 97 (1982).

\smallskip

\item{[2]} A. Chamseddine, Phys. Rev. {\bf B24}, 3065 (1981); L. Castellani,
P. Fr\`e, F. Pilch and P. van Nieuwenhuizen, Ann. of Phys. {\bf 146}, 35 
(1983).

\smallskip

\item{[3]} M.B. Green and J.H. Schwarz, Phys. Lett. {\bf 149 B}, 117 (1984).

\smallskip

\item{[4]} A. Salam and E. Sezgin, Phys. Scr. {\bf 32}, 283 (1985).

\item{} S.J. Gates and H. Nishino, Phys. Lett. {\bf 157B}, 157 (1985).

\smallskip

\item{[5]} A. Strominger, Nucl. Phys. {\bf B343}, 167 (1990); Nucl. Phys.
{\bf B353}, 565 (1991).

\smallskip

\item{[6]} S. Callan, J. Harvey and A. Strominger, Nucl. Phys. {\bf B359},
611 (1991); {\bf B367}, 60 (1991).

\smallskip

\item{[7]} M.J. Duff and J.X. Lu Phys. Rev. Lett. {\bf 66}, 1402 (1991).

\item{} M.J. Duff, R.R. Khuri and J.X. Lu, Nucl. Phys. {\bf B377}, 281 
(1992).

\smallskip

\item{[8]} M.J. Duff, Class Quantum Grav. {\bf 5}, 189 (1988).

\smallskip

\item{[9]} M.J. Duff and J.X. Lu, Nucl. Phys. {\bf B354}, 129 (1991).

\smallskip

\item{[10]} M.J. Duff and J.X. Lu, Nucl. Phys. {\bf B357}, 534 (1991);
Class Quantum Grav. {\bf 9}, 1 (1991); Nucl. Phys. {\bf 354}, 141 (1991);
M.J. Duff and R. Minasian, Nucl. Phys. {\bf B436}, 507 (1995).

\smallskip

\item{[11]} J. Schwarz and A. Sen, Phys. Lett. {\bf B12}, 105 (1993).

\smallskip

\item{[12]} M.J. Duff, R.R. Khuri and J.X. Lu, Phys. Rep. {\bf 259}, 213 
(1995).

\smallskip

\item{[13]} J. Dixon, M.J. Duff and J. Plefka, Phys. Rev. Lett. {\bf 69},
3009 (1992).

\smallskip

\item{[14]} C.M. Hull, hep--th/9512181.

\smallskip

\item{[15]} M.J. Duff, R. Minasian and E. Witten, hep--th/9601036.

\smallskip

\item{[16]} R. Percacci and E. Sezgin, Mod. Phys. Lett. {\bf A10}, 441 
(1995).

\smallskip

\item{[17]} E. Bergshoeff, P.K. Townsend and E. Sezgin, Phys. Lett.
{\bf B189}, 75 (1987).

\smallskip

\item{[18]} J.M. Izquierdo and P.K. Townsend, Nucl. Phys. {\bf B414}, 93 
(1994); J. Blum and J.A.  Harvey, Nucl. Phys. {\bf B416}, 119 (1994).

\smallskip

\item{[19]} A. Candiello, K. Lechner and M. Tonin, Nucl. Phys. {\bf B438}, 
67 (1975).

\smallskip

\item{[20]} K. Lechner and M. Tonin, DFPD/95/TH/63.

\smallskip

\item{[21]}  P.B. Wiegmann, Nucl. Phys. {\bf B323}, 330 (1989).

\smallskip

\item{[22]} J. Honerkamp, Nucl. Phys. {\bf B36}, 130 (1972);
\item{} S. Mukhi, Nucl. Phys. {\bf B264}, 640 (1986);
\item{} M.T. Grisaru and D. Zanon, nucl. Phys. {\bf B310}, 57 (1988);
\item{} J. Atick and A. Dhar, Nucl. Phys. {\bf B284}, 131 (1987).

\smallskip

\item{[23]} L. Alvarez--Gaum\`e  and P. Ginsparg, Ann. Phys. {\bf 161}, 423 
(1985).

\smallskip 

\item{[24]} R.I. Nepomechie, Phys. Rev. {\bf D31}, 1921 (1984);
\item{} C. Teitelboim, Phys. Lett. {\bf B167}, 69 (1986).

\smallskip 

\item{[25]} J. Polchinski, S. Chaudhuri and V. Johnson, hep--th/9602052.
\item{[26]} P. Horava and E. Witten, hep--th/9510209;
\item{} E. Witten, hep--th/9512219.

\smallskip 

\item{[27]} A. Candiello and K. Lechner, Nucl. Phys. {\bf B412}, 479
(1994);
\item{} K. Lechner and M. Tonin, Phys. Lett. {\bf B366}, 149 (1996).

\bye